# Study of Nematic Isotropic Transition

Enakshi Guru
Sarojini Naidu College for Women, Kolkata, India

High accuracy Monte Carlo study has been performed in the modified planar Lebwohl Lasher model containing $P_2$ and $P_4$ interaction, where $P_2$ and $P_4$ are second and fourth order Legendre polynomial having three-dimensional spin. Weakly First order Nematic to Isotropic phase transition is observed using this model. Extensive Monte Carlo simulations using Wolff algorithm are performed to investigate the average energy, order parameter, ordering susceptibility, entropy and specific heat near phase transition. Transition temperature is also estimated.

Key word: Nematic liquid crystal, Lebwohl Lasher model, Monte Carlo simulation, Wolff algorithm, Nematic isotropic phase transition
* Corresponding author: enakshiguru@gmail.com

## 1.Introduction

The Lebwohl-Lasher model [1] consist of a system of rod like molecule placed in a lattice site of three dimensional cubic lattice with interaction potential given by $U_{ij} = -\epsilon_{ij} P_2 cos\theta_{ij}$, where $\theta_{ij}$ is the angle between the axis of the ith molecule and jth molecule and $\epsilon_{ij}$ is the strength parameter. $\epsilon_{ij}$ is equals to $\epsilon$ for i equals to j otherwise its value is zero. The model explains the behaviour of Nematic–isotropic phase transition. It can also be compared with the Ising and Heisenberg model in the field of magnetism. The basic difference of this model with two dimensional O (3) Heisenberg model for magnetic system is that the direction $\hat{n}$ and $-\hat{n}$ are equivalent, $\hat{n}$ being the director. The nematic P2 model and XY model exhibit continuous transition while O(3) Heisenberg model does not exhibit any transition

Extensive Monte Carlo study suggest that Lebwohl-Lasher model exhibits weakly first order phase transition at dimensionless temperature $T_C$, where $T_C = \frac{k_B T}{\epsilon}$ and $k_B$ is the Boltzmann constant. Number of variations of conventional LL model has been investigated by many authors. In planar Lebwohl Lasher model, lattice dimension (d) is 2 and spin dimension (n) is 3. This model was first investigated by Chiccoli.et.al [2]. They use usual Metropolis's algorithm with the periodic boundary condition ranging from $5 \times 5$ to $80 \times 80$. But from this work it was not clear that whether the system exhibits phase transition or not. Kunz and Zumbach [3] execute MC study using cluster algorithm of Wolff [4]. They study $256 \times 256$ lattice. They observed continuous transition at Tc=0.358. They found the strong topological phase transition and correlation length $\xi$ and susceptibility diverges at transition temperature. They also found a casp in specific heat curve.

Despite of detailed study by Kunz and Zumbach, we [5] felt necessary to carry out further investigation with the model. We increase the accuracy of Monte Carlo simulation using cluster algorithm of Wolff at large number of temperatures near $T_C$. We carry out finite size scaling using the specific heat, order parameter susceptibility and autocorrelation time data. The system exhibits second order phase transition at around $T_C$=0.55, which resembles with Kunz Zumbach observation. Recently S. Shabnam et al [6] studied the same model and observed that the system possesses quasi long-range order (QLRO), but they cannot give any conclusion about the order of transition. Another review was done by G.Skacej et.al at his recent work [7].

Zannoni [8] proposed that beside using $P_2 cos\theta$, higher rank interaction like $P_M cos\theta$ with M=4,6 can be introduced. They concluded that the nature of transition will moves towards first order where at transition order parameter and entropy increases. A. Pal et.al [9] uses pure $P_4$ term in the potential and observed first order nematic isotropic transition Chiccoli [10] used $P_4$ interaction term in his observation. Heat capacity is insensitive to the system size as observed by Chicolli et al. If the system exhibit true phase transition heat capacity should sharpened systematically. Romano [11] gives $P_6$ term in his potential. Both of their results agree with the observation of Zannoni [7]. Zhang etal and Priezev [12] et al include P4 term in LL model and observed a weakly 1st order transition.

The P4 system exhibits strongly first order transition where P2 model exhibit continuous transition. After a decade, the model is revisited by me in slightly different approach. Non negligible higher rank contribution is expected for real molecules. So, it is considered to add even rank term ($P_4$) term to the potential and the contribution of $P_2$ and $P_4$ are equally shared. n this paper, the mixed effect of second and fourth rank term in

1:1 ration in the potential of a nematic is studied systematically. The objective of this paper is to study the nature of transition. At least ten temperatures are investigated around transition temperature for L=50, 70 and 100 lattice sizes. In Figure1 potential of the used system is plotted with respect to $\theta$ and for varying strength of fourth rank contribution.

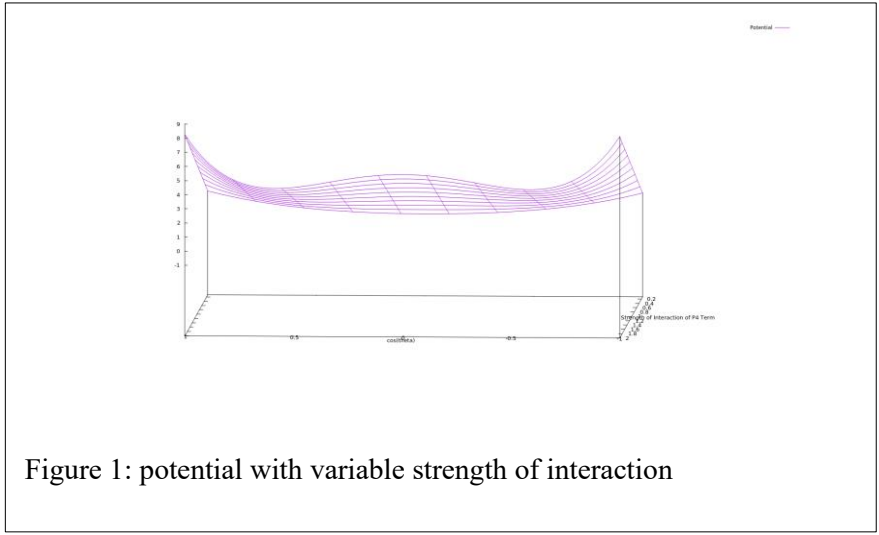

Figure 1: potential with variable strength of interaction

The simulation is performed in square lattice using cluster algorithm of Wolff. Periodic boundary condition is also used. Energy, second rank orientational order parameter is calculated. Specific heat and order parameter susceptibility is calculated for modified model by fluctuation technique. Average cluster size using Wolff algorithm is also noted with temperature. Auto correlation time **[13]** is measured for different lattice sizes. Pair correlation time is also calculated.

## 2. Computational Method
In this section we will discuss different mathematical treatment that are used in this paper to study this model.
### 2.1 Wolff Algorithm for Monte Carlo method
In context with liquid crystalline system, we consider the Hamiltonian of modified Lebwohl-Lasher model is

$$H = -\epsilon \sum_{\langle x,y \rangle} \frac{3\{(\vec{\sigma_x},\vec{\sigma_y})^2 - 1\}}{2} + \left[\frac{3\{(\vec{\sigma_x},\vec{\sigma_y})^2 - 30(\vec{\sigma_x},\vec{\sigma_y}) + 3\}}{8}\right]$$

We choose a random vector $\vec{\sigma_x}$ in a random lattice site at first point. We flip $\vec{\sigma_x}$ by
$\vec{\sigma_x'} = \hat{R}(x)\vec{\sigma_x} = \vec{\sigma_x} - 2(\vec{\sigma_x},\vec{r})\vec{r}$.
We link all nearest neighbour y connecting x. The bond $\langle x, y \rangle$ is activated by the probability
$p(x,y) = 1 - exp\left(min\{0, \beta\vec{\sigma_x}, [1 - \hat{R}(x)]\vec{\sigma_y}\}\right)$.
If the property is satisfied $\vec{\sigma_y}$ is flipped. y is marked and adjoined in the cluster. The process continues iteratively in the same way for all bonds leading to unmarked neighbours of adjoined sites until the process stops.

### 2.2 Determination of Auto correlation time

The autocorrelation function gives a measure of the correlation of a parameter at two different times, one an interval t later than t other. Let {A(t)} be a real -valued stationary stochastic process with mean $D \equiv \langle A(t) \rangle$. $D = \bar{A} = \frac{1}{n}\sum_{i=1}^{n} A_i$ In this paper we use A(t) as order parameter and energy. The unnormalized autocorrelation function is given by $C(t) \equiv \langle A_i A_{i+t} \rangle - D^2$. The auto-correlation time drops from a significant non-zero value at a short time t towards zero at very long times. The normalized autocorrelation function is given by

$$\rho(t) \equiv C(t)/C(0)$$

The typical time scale at which the ratio falls off is a measure of the correlation time $\tau$ of the simulation. If we integrate A(t) over time, it will give nonzero value if on the average the fluctuations are correlated, or it will be zero if they are not. The natural estimator of C(t) is

$\hat{C}(t) = \frac{1}{n-|t|}\Sigma_{A_i - \mu}(A_{i+|t|} - D)$ where n is the sample size and $\mu = \langle A(t) \rangle$ The natural estimator $\rho(t)$ is
$\hat{\rho}(t) \equiv \frac{\hat{C}(t)}{\hat{C}(0)}$

Integrated auto-correlation time is given by

$$\tau_f = \frac{1}{2}\Sigma_{t=-(n-1)}^{(n-1)} \lambda(t)\hat{\rho}(t)$$

$\lambda(t) = 1$, for $|t| \leq \tau$. But $\lambda(t) = 0$, for $|t| \gg \tau$ (as noise is present). This retains most of the signal and discard noise.

## 2.3 Determination of thermodynamic quantities by fluctuation

Potential used in this model is given by,

$$E = -\epsilon \left[\left(\frac{3\cos^2\theta - 1}{2}\right) + C\left(\frac{35\cos^2\theta - 30\cos\theta + 3}{8}\right)\right]$$

Where C is the strength parameter. In this model, we choose C=1 and strength of interaction $\epsilon = 1$. The $P_2$ and $P_4$ are potential contribute to the Hamiltonian equally (50%).

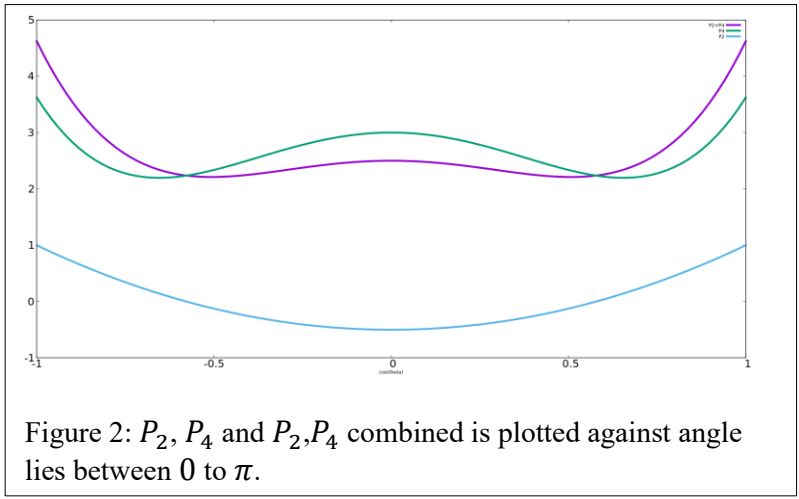

Figure 2: $P_2$, $P_4$ and $P_2, P_4$ combined is plotted against angle lies between 0 to $\pi$.

The mean field potential which is proportional to $P_M \cos\theta$, possesses $\frac{M+2}{2}$ minima for $\theta$ lying between $[0, \pi]$. For interaction potential $P_M(\cos\theta)$ where M =2, 4, 6 …even number, the minima at $\theta = 0$ and $\theta = \pi$ becomes narrower and steeper with increasing value of M. So, if we increase the rank of interaction there may be a chance of a particle to be trapped inside the minima rather than inclined towards the director. The $P_4$ model has three minima within $[0, \pi]$ and magnitude of minima is steeper than mixture of $P_2, P_4$.

The average energy per particle of the system is given by $\langle E \rangle = \frac{1}{N}\Sigma_{i=1}^{N} E_i$, N being the total number of particles in the system and the reduced specific heat per particle is given by $C_v = \frac{1}{L^2}\frac{d}{dT}\langle E \rangle = \frac{d}{dT}\langle E \rangle$, where the angular brackets represent the ensemble average and T is the dimensionless temperature. Specific heat can also be evaluated from the fluctuation of energy

$$C_v = \frac{1}{T^2}(\langle E^2 \rangle - \langle E \rangle^2).$$

The usual long range order parameter is defined as $\langle P_2 \rangle = \frac{1}{2}\langle 3\cos^2\theta - 1 \rangle$, which is obtained by maximizing $\langle P_2 \rangle$. The order parameter susceptibility defined as the fluctuation of order parameter

$$\chi_0 = \frac{1}{T}(\langle P_2^2 \rangle - \langle P_2 \rangle^2)$$

The second rank pair correlation function is defined as $G_2(r) = \langle P_2(\cos\alpha_{ij}) \rangle_r$ where $\alpha_{ij}$ is the angle between two spins $i$ and $j$ separated by a distance r.

## 4. Result and discussion

We use planar $P_2P_4$ model having nearest neighbour interaction, where $P_2$ and $P_4$ are the second and fourth order Legendre Polynomial. The model describes a two dimensional nematic having three-dimensional spin. Monte Carlo simulation were performed on planar model of lattice size L=50, 70, 90 and 100 using the cluster algorithm of Wolff [4] instead of spin flip algorithm of Metropolis. Ergodicity and detailed balance condition are maintained in this technique. Periodic boundary condition is used. For each lattice size ten to eleven temperatures close to the transition were used. Three million of Monte Carlo (MC) steps are generated to reach in equilibrium and averages of thermodynamic quantities are measured with 3 million.

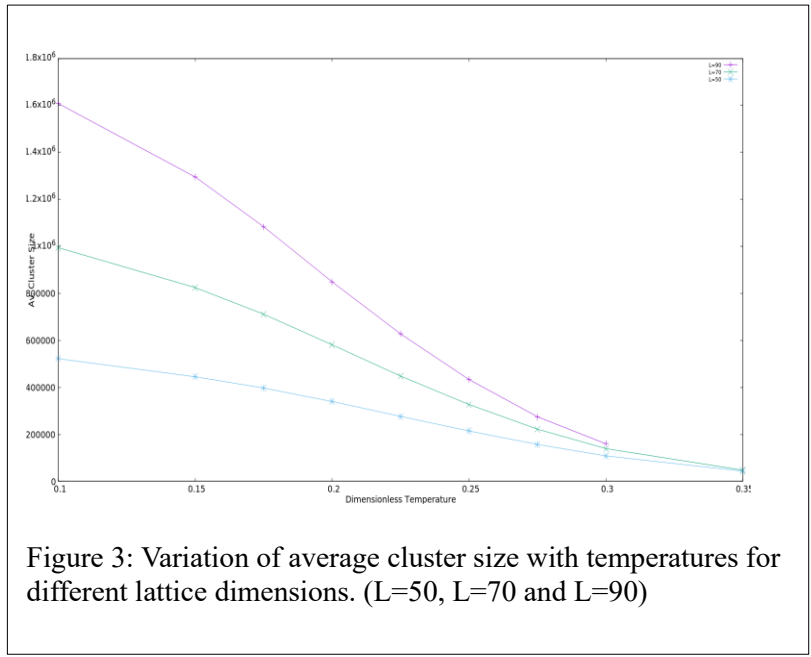

Figure 3: Variation of average cluster size with temperatures for different lattice dimensions. (L=50, L=70 and L=90)

It is seen from the Figure 3 that average size of Wolff cluster decreases with temperature. It is plotted for L=50, L=70 and L=100 lattice size. The nature is same for all lattice sizes. For higher lattice size cluster size increases.

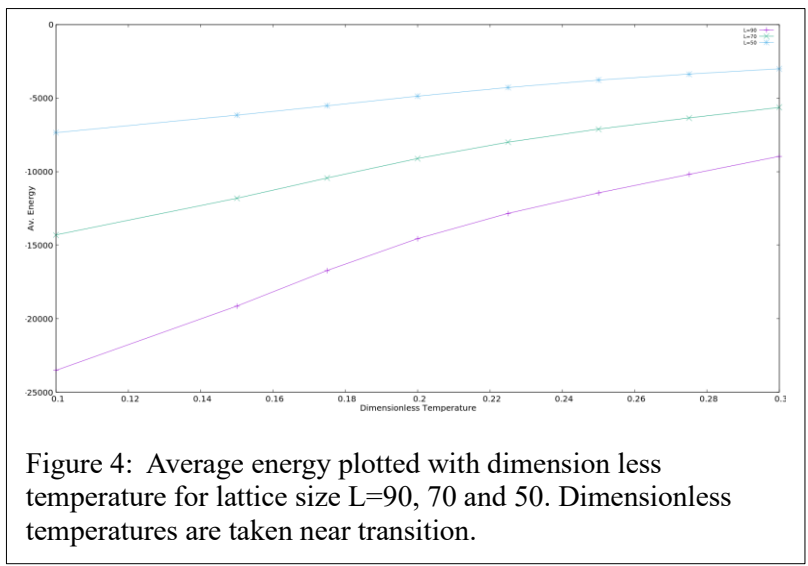

Figure 4: Average energy plotted with dimension less temperature for lattice size L=90, 70 and 50. Dimensionless temperatures are taken near transition.

The average energy is plotted with temperature. It is seen from the graph that there is increase in energy between $T^* = 0.1$ to $T^* = 0.3$ In case of pure $P_4$ model, there was a sharp transition [6] of average energy which indicates the first order transition. In $P_2$ model, increase in energy is continuous [5].

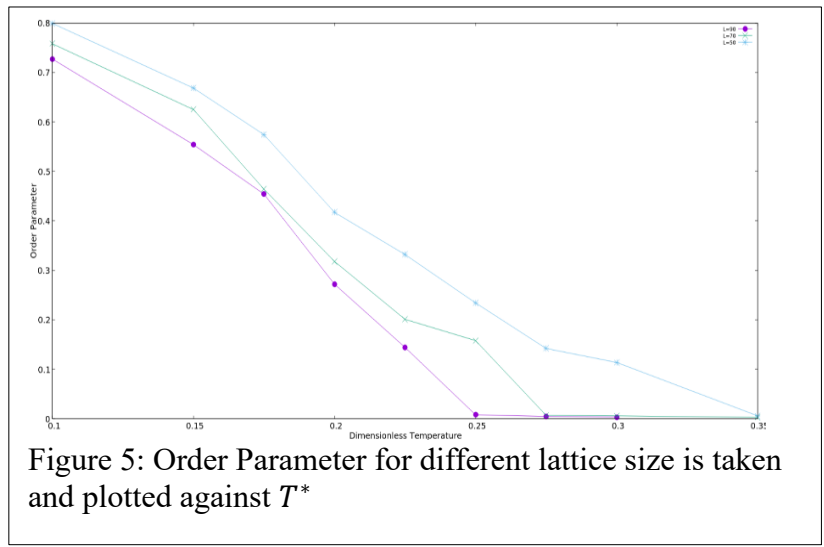

Figure 5: Order Parameter for different lattice size is taken and plotted against $T^*$

The orientational order parameter is plotted near transition and rapid fall of order parameter from nematic phase to isotropic phase is clearly seen. From the graph it can be concluded that transition takes place in between the dimensionless temperature 0.15 to 0.25. The nature of the graph shows the evidence of first order transition.

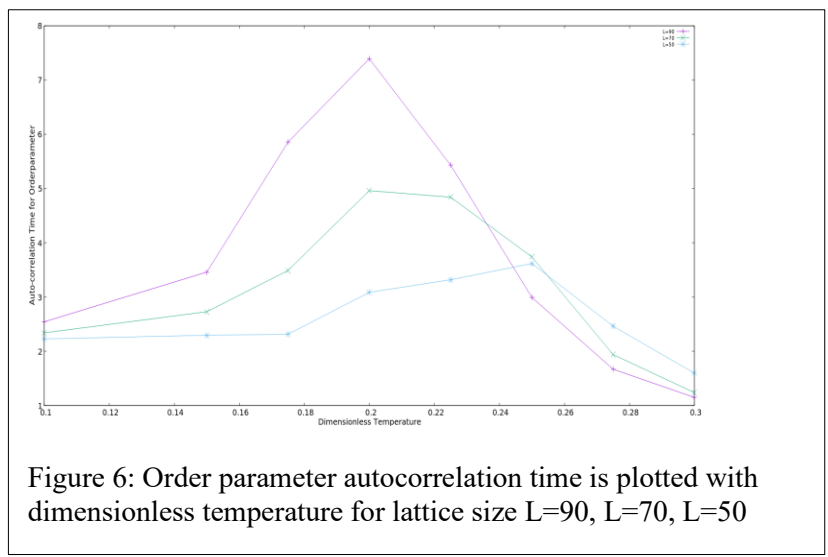

Figure 6: Order parameter autocorrelation time is plotted with dimensionless temperature for lattice size L=90, L=70, L=50

The autocorrelation of energy and order parameter are observed. Autocorrelation of order parameter has definite peak, but autocorrelation of energy does not show any peak. In Figure 6 the autocorrelation time for order parameter is plotted for sizes L=50, 70, 90. The peak shifts to lower temperature side as the lattice size increases. In pure $P_2$ model we also get the peak in auto correlation time of order parameter and concluded Wolff cluster algorithm is more efficient than Metropolis as the critical exponent is low for this technique [5]. In Table1 auto correlation of energy and specific heat at different temperatures are given for lattice sizes L=50, 70 and 90 around transition temperature. The data shows that there is no peak for energy auto-correlation time. Specific heat also shows continuous decrease in values. Specific heat at constant volume is also noted from fluctuation. No peak but rapid decrease in $C_V$ is observed from fluctuation data but if the derivative of average energy is taken, peak is observed. (Figure 7)

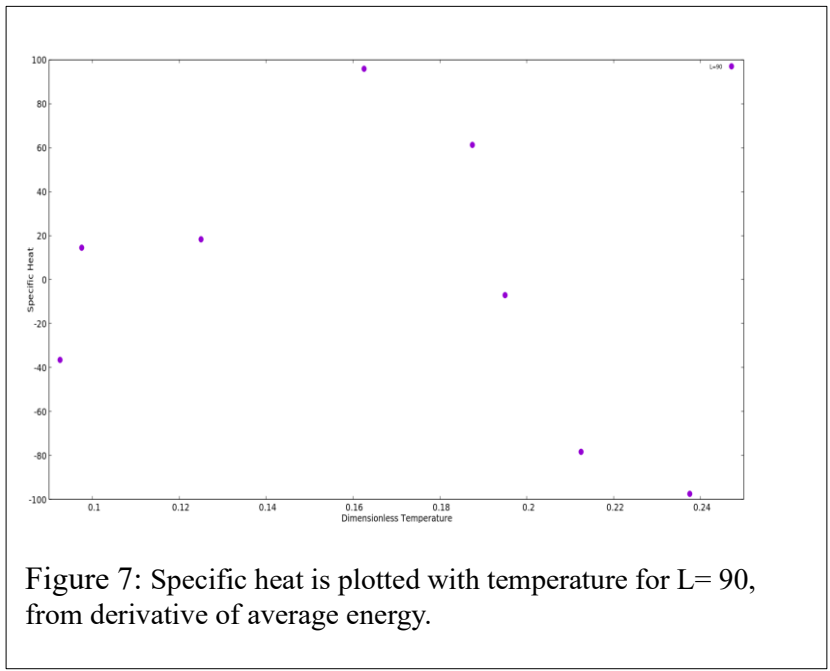

Figure 7: Specific heat is plotted with temperature for L= 90, from derivative of average energy.

System entropy is calculated for ten to eleven temperatures around transition for L=90, 70 and 50. It increases with lattice size but for all lattice sizes the variation of entropy with temperature is nearly equal

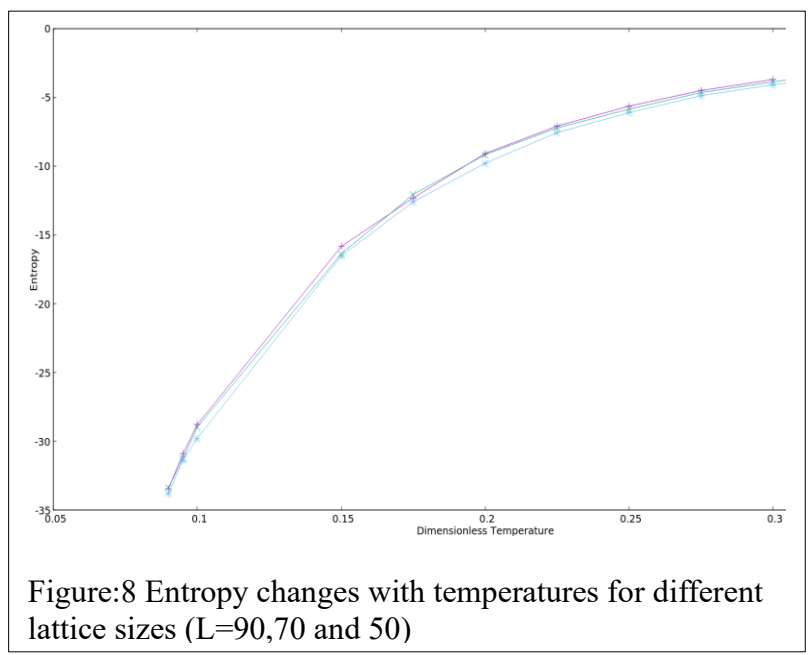

Figure:8 Entropy changes with temperatures for different lattice sizes (L=90,70 and 50)

Ordering susceptibility is plotted with temperature for varying lattice sizes in Figure 9. It shows a peak, whereas in our earlier wok with $P_2$ model [5] the peak is sharper. Specific heat from fluctuation is also plotted for different temperatures. Here also flatten peak is observed.

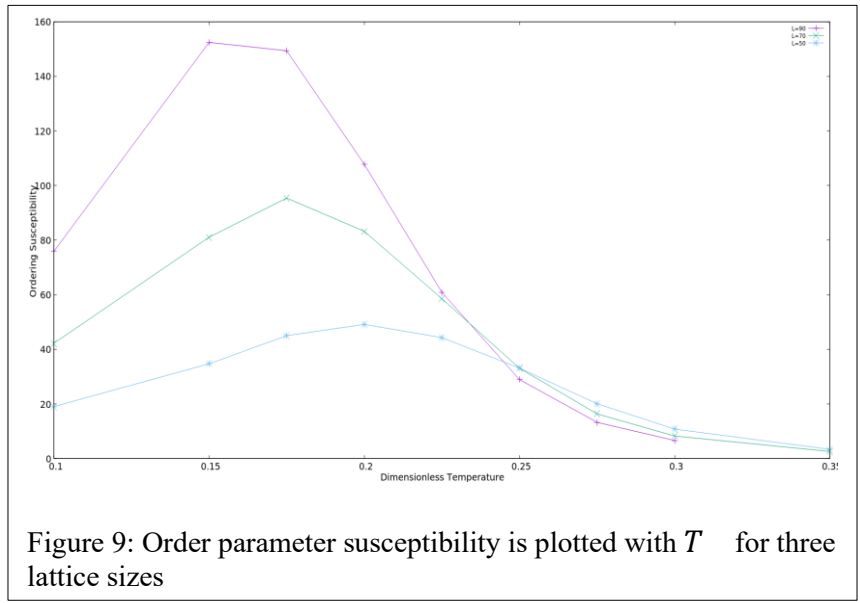

Figure 9: Order parameter susceptibility is plotted with $T$ for three lattice sizes

For plotting pair correlation, the range of r has been chosen as half of the lattice size. For the system exhibiting true long-range order, the correlation function decays to a plateau for large r and the limiting value is close to $\langle P_2 \rangle^2$. If the system exhibits quasi long range order the correlation decays like power law as it happens in XY model. In disordered system it dies of exponentially. In Figure 10 pair correlation function for lattice dimension L=70 the angular correlation plot is fitted in Figure 11 and 12. For temperature below Tc (T*=0.1) the angular correlation fits with power law $f(x) = a + bx^{-p}$, where a=0.5677, b=0.2964, c=0.4574. This implies the system exhibits quasi long-range order (QLRO). For the temperature lower than transition temperature $T^* = 0.1$, $G_2(r)$ could be fitted accurately with a power law decay to a plateau.

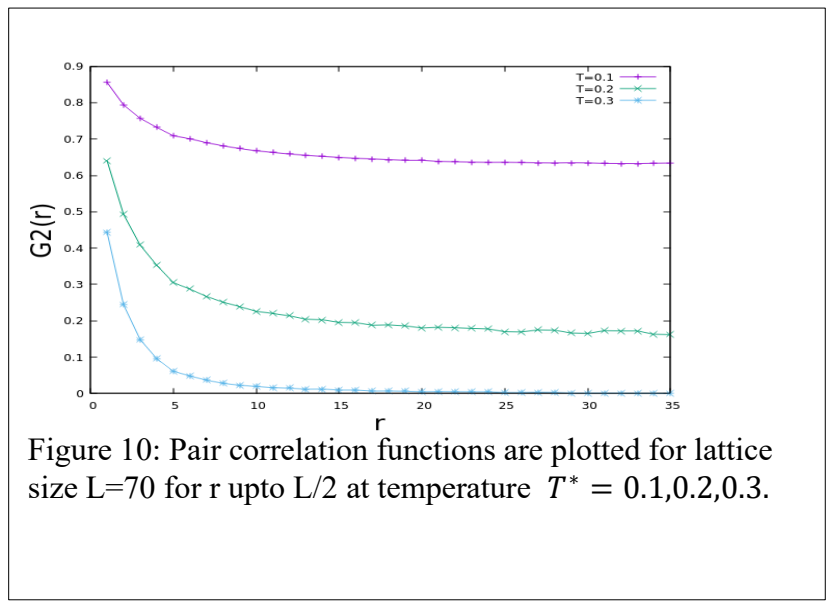

Figure 10: Pair correlation functions are plotted for lattice size L=70 for r upto L/2 at temperature $T^* = 0.1, 0.2, 0.3$.

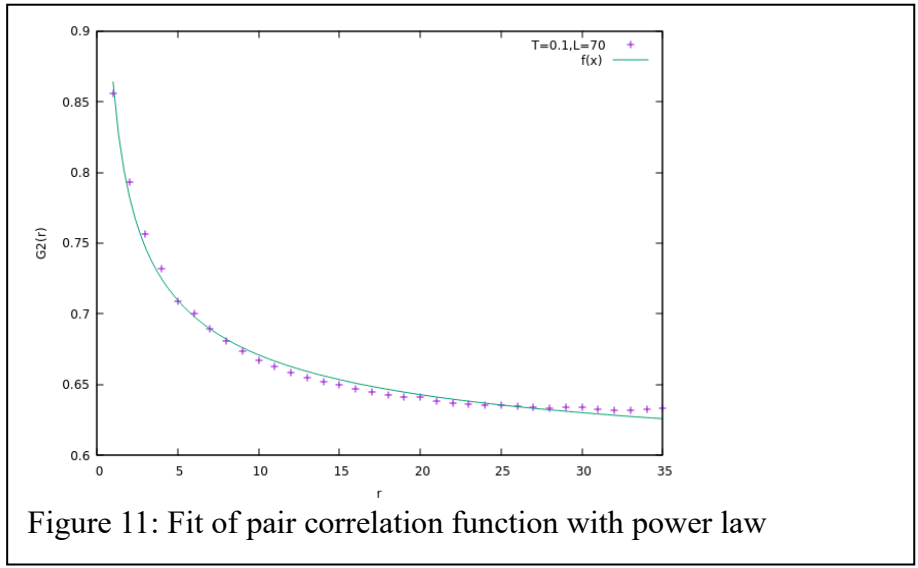

Figure 11: Fit of pair correlation function with power law

exponentially fit agrees reasonably well at temperature above Tc (T*=0.3). For more higher temperature may be exact fit is possible. The fit parameter c is equals to 1.31826. It implies the system reaches to disordered state. It clearly signifies that system goes to QLRO to ordered state transition.

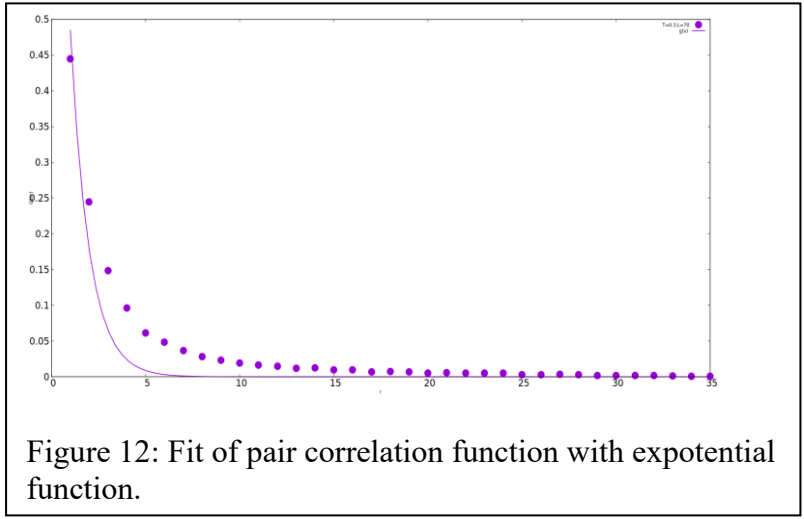

Figure 12: Fit of pair correlation function with expotential function.

**Conclusion**

The objective of the paper is to study the effect of P2 P4 potential in Nematic Isotropic transition. Planar Lebwohl Lasher model is used with 3 dimensional spin interacting with P2 P4 potential  Extensive Monte Carlo simulation is used to study at least 10 temperature around transition temperature. Monte carlo study of P2P4 model reveals that the sytem exhibits QLRO to disordered transition The nature of transition is very weakly first order. The transition temperature is around 0.175. The range is within 0.15 to 0.2

| $T$ | L=50 | | L=70 | | L=90 | |
|---|---|---|---|---|---|---|
| | $\tau_{energy}$ | $c_V$ | $\tau_{energy}$ | $c_V$ | $\tau_{energy}$ | $c_V$ |
| 0.09 | 3.66 | 427.99 | 4.61 | 788.85 | 4.72 | 935.68 |
| 0.095 | 3.83 | 414.33 | 3.91 | 631.39 | 4.52 | 916.81 |
| 0.1 | 3.60 | 398.09 | 3.98 | 623.92 | 4.73 | 904.71 |
| 0.15 | 3.90 | 331.35 | 4.62 | 534.97 | 5.58 | 780.138 |
| 0.175 | 3.83 | 282.80 | 5.10 | 397.86 | 6.70 | 461.97 |
| 0.2 | 4.68 | 195.96 | 5.34 | 213.58 | 6.10 | 201.90 |
| 0.225 | 4.49 | 117.72 | 5.08 | 42.53 | 5.86 | 99.68 |
| 0.25 | 4.67 | 70.47 | 5.87 | 63.32 | 8.15 | 58.93 |
| 0.275 | 5.33 | 45.70 | 9.28 | 42.53 | 15.85 | 40.70 |
| 0.3 | 7.61 | 33.33 | 16.43 | 31.44 | 27.67 | 30.12 |
| 0.35 | 23.25 | 21.53 | 38.8 | 18.51 | - | - |

Table 1: Autocorrelation of energy and specific heat are given for eleven temperatures around transition for three different lattice sizes